\documentclass[apj,iop,twocolumn]{emulateapj}
\usepackage{amsmath}
\usepackage{float}
\usepackage{nicefrac}
\usepackage{rotating}
\usepackage{rotfloat}
\usepackage[none]{hyphenat}
\usepackage{enumitem}
\usepackage{afterpage}

\def\aj{AJ}

\def\apj{ApJ}

\def\apss{Ap\&SS}
\def\aap{A\&A}

\def\aaps{A\&AS}

\def\mnras{MNRAS}

\def\pasa{Publ.~Astron.~Soc.~Australia}

\def\rmxaa{RMxAA}

\def\arcsec{\hbox{$^{\prime\prime}$}}

\shorttitle{Collimated Bipolar Outflows in Th 2-A}
\shortauthors{Danehkar}

\begin{document}

\title{Discovery of collimated bipolar outflows in the planetary nebula Th\,2-A}

\author{A. Danehkar\altaffilmark{1}}
\affil{Department of Physics and Astronomy, Macquarie University, Sydney, NSW 2109, Australia}
\email{ashkbiz.danehkar@cfa.harvard.edu}

\altaffiltext{1}{Present address: Center for Astrophysics, 60 Garden Street, Cambridge, MA 02138, USA}

\begin{abstract}
We present a comprehensive set of  spatially resolved, integral field spectroscopic mapping of the Wolf--Rayet planetary nebula Th\,2-A, obtained using the Wide Field Spectrograph on the Australian National University 2.3-m telescope. Velocity-resolved H$\alpha$ channel maps with a resolution of $20$\,km\,s$^{-1}$ allow us to identify different kinematic components within the nebula. This information is used to develop a three-dimensional morpho-kinematic model of the nebula using the interactive kinematic modeling tool \textsc{shape}. These results suggest that Th\,2-A has a thick toroidal shell with an expansion velocity of $40\pm10$\,km\,s$^{-1}$, and a thin prolate ellipsoid with collimated bipolar outflows toward its axis reaching velocities in the range of $70$--110\,km\,s$^{-1}$, with respect to the central star. The relationship between its morpho-kinematic structure and  \textit{peculiar} [WO]-type stellar characteristics deserves further investigation.
\end{abstract}


\section{Introduction}
\label{th2a:sec:introduction}

The planetary nebula (PN) Th\,2-A (\,=\,PN G306.4$-$00.6) was classified by \citet{Kromov1968} under the ring-shaped objects. This morphology is also visible in the H$\alpha$ image taken by \citet{Gorny1999}. The abundance analysis of Th\,2-A yielded He/H~=~0.09 and N/O~=~0.40  \citep{Kingsburgh1994}, and He/H~=~0.13 and N/O~=~0.49 \citep{Henry2004}. The oxygen abundance was found to be close to the solar metallicity \citep[O/H~=~$4.67\times10^{-4}$;][]{Kingsburgh1994}. The nebular He\,{\sc ii} $I$(4686$)/I$(H$\beta)=0.50$, measured by \citet{Kingsburgh1994} corresponds to a relatively high-excitation nebula. This PN has a moderate density of  $N_{\rm e}$(S\,{\sc ii})~=~1220\,cm$^{-3}$ \citep{Kingsburgh1994}, which could be associated with a relatively hot central star (CSPN) based on the evolutionary radiation-hydrodynamic models \citep{Perinotto2004}.

This PN is indeed ionized by a very hot star classified as [WO3]${}_{\rm pec}$ \citep{Weidmann2008}. Previously, \citet{Acker2003} defined the [WO]${}_{\rm pec}$ subclass as those with a \textit{peculiar} C\,{\sc iv} 5805\,{\AA} full width at half maximum (FWHM), much wider than typical [WO]. Similarly, \citet{Weidmann2008} identified the CSPN Th\,2-A as type [WO]${}_{\rm pec}$, belonging to those with \textit{peculiar} C\,{\sc iv}-5801/12 doublet. According to the transformation given by \citet{Dopita1991}, the strength of He\,{\sc ii} $\lambda$4686 relative to H$\beta$ describes a stellar temperature of $T_{\rm eff}=158$\,kK. However, the He\,{\sc ii} and [O\,{\sc iii}] emission lines can also correspond to $T_{\rm eff}=291$\,kK based on the method introduced by \citet{Reid2010}. 

Th\,2-A was found to contain a possible binary system \citep{Ciardullo1999}. One component of the binary is a CSPN, and another component is a visual late-type star separated by $1\farcs 4$ \citep{Weidmann2008}. However, the resolved companion could be a possible superposition. Previously, \citet{Kerber2003} described Th\,2-A as PN with a well defined photometric center. Although there are no detailed studies on the visual companion spectra, \citet{Weidmann2008} identified broad and intense C\,{\sc iv} and O\,{\sc iv} emission lines in the CSPN spectra, which made it one of the rare [WO]${}_{\rm pec}$ stars. Notwithstanding, if the nearby visual star is indeed a binary companion of the CSPN, it unlikely contributes to the formation of the fast stellar winds identified in its spectra since its separation is 3500 AU at the distance of 2.5\,kpc determined by \citet[][]{Stanghellini2010}.

In this paper we present integral field spectroscopic mapping of Th\,2-A, from which we derive a morpho-kinematic model of this object. Currently, there are no previous morpho-kinematic studies on this object in the literature. In Section~\ref{th2a:sec:observations}, we describe the details of the observation. In Section~\ref{th2a:sec:results}, we discuss the results. Section~\ref{th2a:sec:models} presents the morpho-kinematic model of Th\,2-A. Finally, we finish with our conclusions in Section~\ref{th2a:sec:conclusions}.

\section{Observations}
\label{th2a:sec:observations}

Integral field spectroscopic observations of Th\,2-A were obtained using the Wide Field Spectrograph \citep[WiFeS;][]{Dopita2007,Dopita2010} on the Australian National University (ANU)\,2.3-m Advanced Technology Telescope (ATT) at the Siding Spring Observatory on 2010 April 20 under program number 1100147 (PI: Q.A. Parker). WiFeS is an image-slicing Integral Field Unit (IFU) feeding a double-beam spectrograph. WiFeS samples $0\farcs5$ along each of twenty five $38\arcsec  \times 1\arcsec $  slitlets, which provides a field of view of $25\arcsec \times 38\arcsec$ and a spatial resolution element of  $1\farcs0\times0\farcs5$. The output is optimized to fit the $4096 \times 4096$ pixel format of the CCD detectors. Each slitlet is designed to project to 2 pixels on the detector. This yields a reconstructed point spread function with a FWHM of about 2 arcsec. The spectrograph uses volume phase holographic gratings to provide a spectral resolution of either $R \sim 3000$ or $R \sim 7000$. We used the spectral resolution  of $R\sim 7000$, resulting in a linear wavelength dispersion per pixel of $0.45$\,{\AA} for the red spectrum (5222--7070\,{\AA}). This spectral resolution yields a resolution of $\sim 20$\,km\,s${}^{-1}$. 

We reduced the data using the \textsc{iraf} pipeline \textsf{wifes} to correct bias, correct pixel-to-pixel sensitivity using dome flat-field frames, calibrate spectra based on arc lamp exposures, calibrate space based on wire frames, correct differential atmospheric refraction, remove cosmic rays, and calibrate data to the absolute flux unit based on spectrophotometric standard stars \citep[described in detail by][]{Danehkar2013a,Danehkar2014a}.

\begin{figure}
\begin{center}
\includegraphics[width=2.5in]{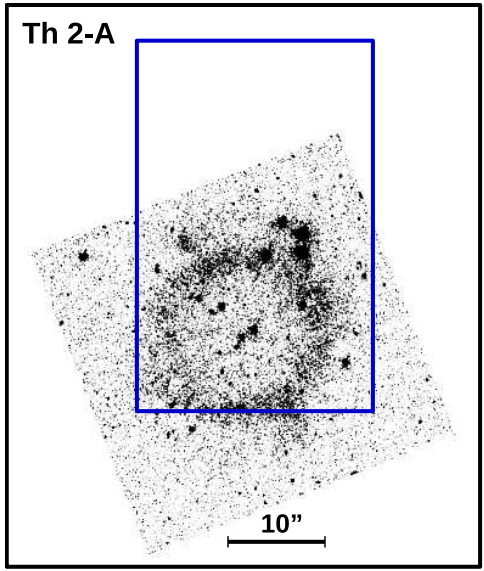}\\
\caption{\textit{HST} image of Th\,2-A on a squared scale taken with the F555W filter and WFPC2 on \textit{HST} (observing program 6119).
The rectangle shows the $25\arcsec \times 38\arcsec$ WiFeS field of view observed using the ANU\,2.3-m telescope at the Siding Spring Observatory in 2010 April. North is up and east is toward the left-hand side.
\label{th2a:hst:fig1}%
}%
\end{center}
\end{figure}

The \textit{Hubble Space Telescope} (\textit{HST}) image of Th\,2-A shown in Fig.\,\ref{th2a:hst:fig1} was retrieved from the \textit{HST} archive. The \textit{HST} image was taken using the non-aberrated Wide Field Planetary Camera 2 (WFPC2) and the F555W (``\textit{V}'') filter with an exposure time of 60\,s  under program 6119 (PI: H. Bond) on 1995 September 23, which has a central wavelength of 5439\,{\AA} and a bandwidth of 1228\,{\AA}. The rectangle in Fig.\,\ref{th2a:hst:fig1} shows the location and area of the WiFeS aperture in the nebula.

\begin{figure*}
\begin{center}
\includegraphics[width=1.70in]{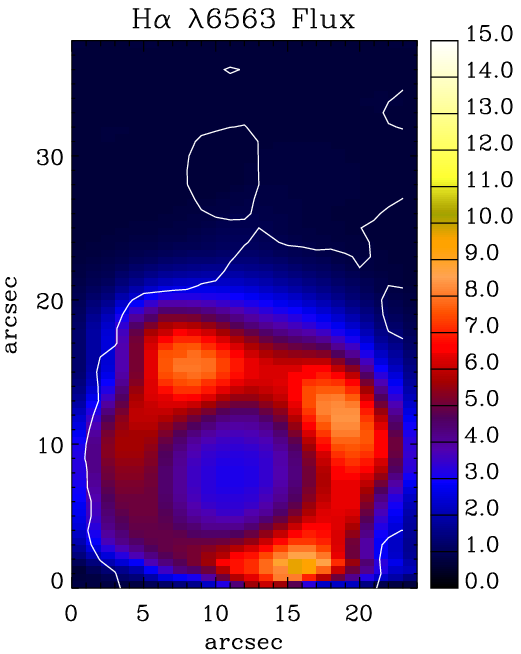}%
\includegraphics[width=1.70in]{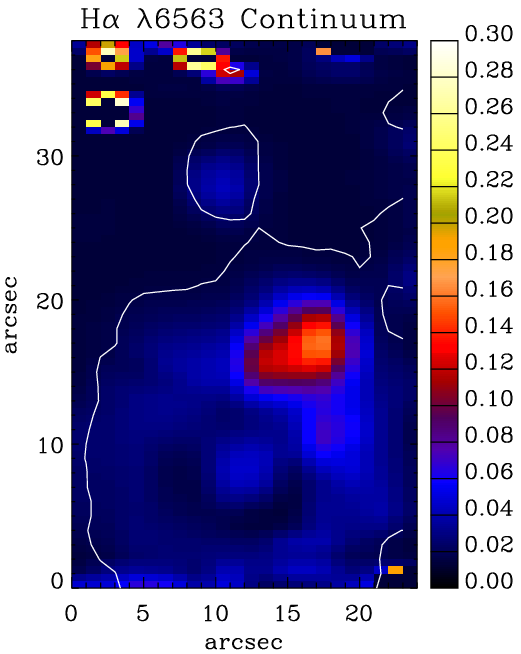}%
\includegraphics[width=1.70in]{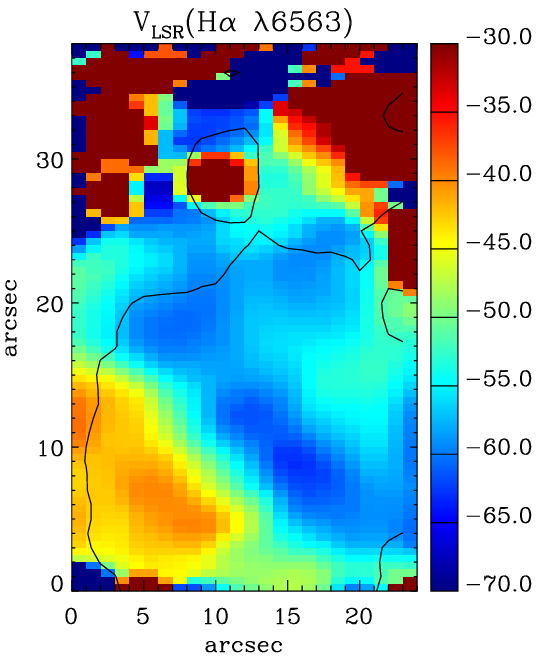}
\includegraphics[width=1.70in]{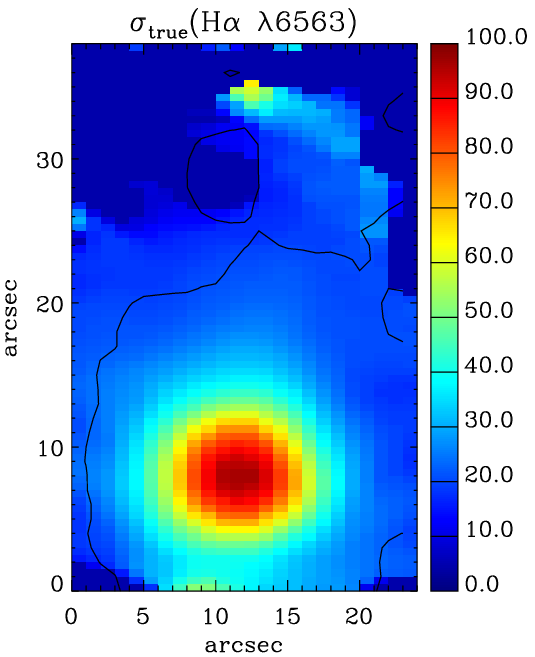}
\caption{Kinematic maps of Th\,2-A in H$\alpha$ $\lambda$6563\,{\AA}. From
left to right: the integrated line flux, continuum flux, radial velocity map (LSR), and velocity dispersion map. 
The color bars of the panels showing flux measurements are in $10^{-15}$~erg\,s${}^{-1}$\,cm${}^{-2}$\,spaxel${}^{-1}$ unit, those showing continuum are in $10^{-15}$~erg\,s${}^{-1}$\,cm${}^{-2}$\,{\AA}${}^{-1}$\,spaxel${}^{-1}$ unit, and those showing velocity measurements are in km\,s${}^{-1}$ unit.  
The white/black contours in the four maps are the narrow-band H$\alpha$ emission in arbitrary unit obtained from the SHS. North is up and east is toward the left-hand side.
\label{th2a:fig2}%
}%
\end{center}
\end{figure*}

\section{Results}
\label{th2a:sec:results}

\begin{figure*}
\begin{center}
\includegraphics[width=6.0in]{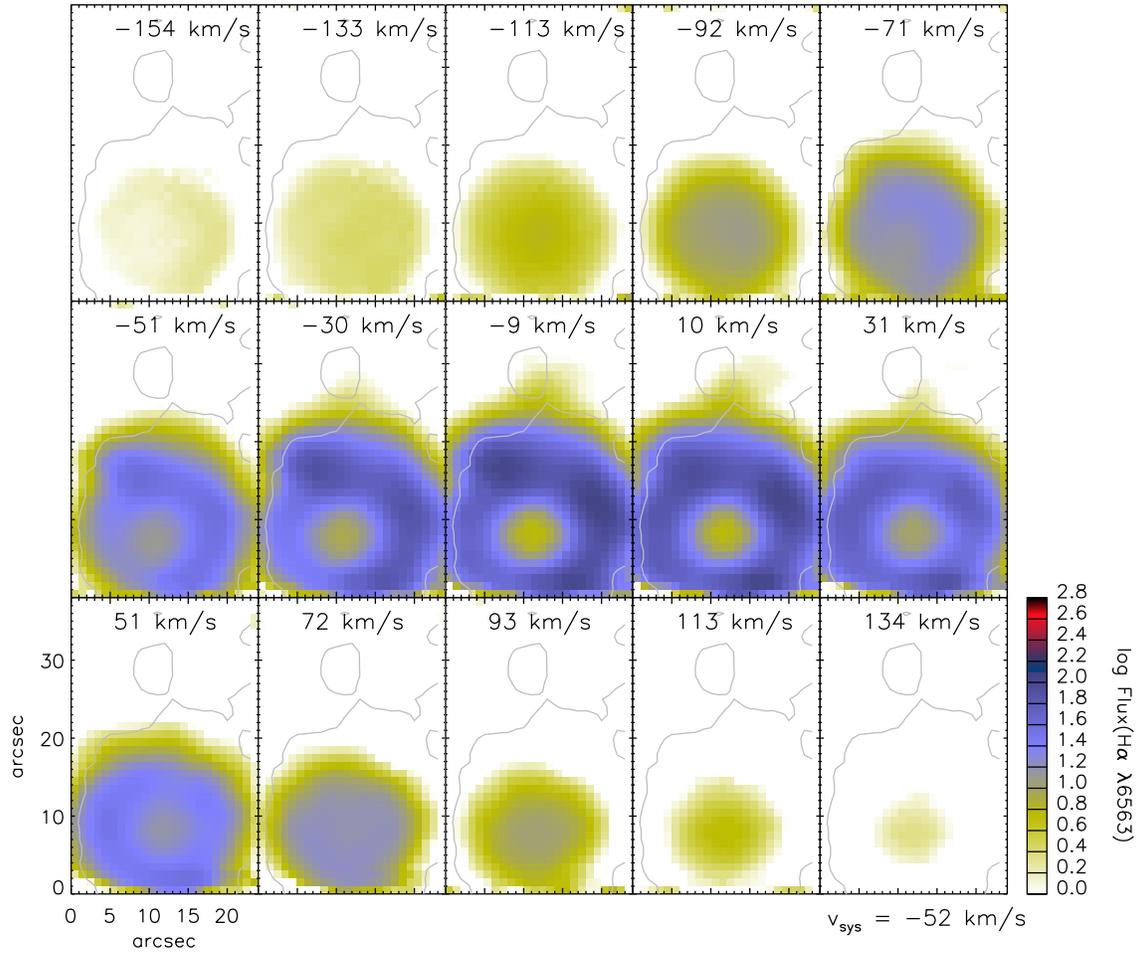}\\
\caption{Velocity slices of Th\,2-A along the H$\alpha$ emission-line profile. The slices have a $\sim 20$\,km\,s$^{-1}$ width, the central velocity is given at the top of each slice, and the LSR systemic velocity is $v_{\rm sys}=-52$\,km\,s$^{-1}$. The color bar shows flux measurements in logarithm of $10^{-15}$~erg\,s${}^{-1}$\,cm${}^{-2}$\,spaxel${}^{-1}$ unit. Velocity channels are in km\,s${}^{-1}$ unit. The contours in the channel maps are the narrow-band H$\alpha$ emission in arbitrary unit obtained from the SHS.  North is up and east is toward the left-hand side. 
\label{th2a:vmap}%
}%
\end{center}
\end{figure*}

In Figure \ref{th2a:fig2}, we present the spatially resolved maps of the flux intensity, continuum, radial velocity and velocity dispersion of H$\alpha$ $\lambda$6563 for Th\,2-A, obtained by fitting a Gaussian curve to each spaxel of the IFU datacube using the \textsc{mpfit} routine \citep{Markwardt2009}. The white/black contour lines in the maps depict the distribution of the H$\alpha$ emission obtained from the SuperCOSMOS H$\alpha$ Sky Survey \citep[SHS;][]{Parker2005}, which can aid us in distinguishing the nebular border. We corrected the radial velocity for the motions of the Earth and Sun at the time of our observation by using the \textsc{iraf} task \textsf{rvcorrect}, that results in the local standard of rest (LSR) radial velocity. We corrected the velocity dispersion for the instrumental width, the thermal broadening and the fine structure broadening. The instrumental width is obtained from the $[$O\,{\sc i}$]\,\lambda$6300 night sky line, and is typically $\sigma_{\rm ins}\approx 18$\,km\,s$^{-1}$ at the chosen spectral resolution of $R\sim7000$. The thermal broadening $\sigma_{\rm th}$ is obtained through the Boltzmann's equation $\sigma_{\rm th}= \sqrt{8.3\,T_e[{\rm kK}]/Z}$~[km\,s$^{-1}$], where $T_e$ is the electron temperature of the nebula and $Z$ is the atomic weight of the atom or ion. The fine structure broadening $\sigma_{\rm fs}$ in the hydrogen recombination lines is typically $\sigma_{\rm fs}\approx3$\,km\,s${}^{-1}$ for H$\alpha$ \citep{Clegg1999}. 

As seen in Figure \ref{th2a:fig2}, the flux map shows a polygonal ring shape with low emission at the central region \citep[also visible in the narrow-band H$\alpha$ image of][]{Gorny1999}. However, the \textit{HST} image shown in Fig. \ref{th2a:hst:fig1} does not show this morphology since it was taken using the wide-band F555W filter. The appearance of a prolate nebula viewed almost pole-on can become  rectangular, when the density distribution along the shell decreases slightly with distance from the equator \citep[see e.g.][]{Akras2012b}. The high values of velocity dispersion seen at the central region could be related to high-velocity point-symmetric outflows toward the axis of a prolate ellipsoid viewed pole-on, which is easily noticeable in the channel maps (see Fig. \ref{th2a:vmap}) and is discussed below (and later in Section~\ref{th2a:sec:models}). 

The expansion velocity ($v_{\rm exp}$) obtained from the half width at half maximum (HWHM) of the H$\alpha$ emission flux integrated over the entire nebula is $V_{\rm HWHM}=40\pm5$\,km\,s$^{-1}$, obtained from the corrected dispersion velocity, i.e. $V_{\rm HWHM}=({8\,{\rm ln (2)}})^{1/2}\sigma_{\rm corr}/2$. The obtained HWHM expansion velocity is higher than the peak-to-peak velocity of $2V_{\rm exp}=35$\,km\,s$^{-1}$ derived from the [O\,{\sc iii}] emission line by \citet{Meatheringham1988}. Note that the [O\,{\sc iii}] emission typically occurs near inner regions of the nebula whose expansion velocity is lower than outer regions. However, measuring the expansion velocity by means of the HWHM method is not very fruitful for more detailed kinematic studies. The radiation-hydrodynamics models by \citet{Schonberner2010} showed that the HWHM velocities of volume-integrated line profiles always underestimate the true expansion velocity.  The HWHM method is suitable for slowly expanding objects, but it does not reflects real expansion velocities of large spatially resolved objects. 

Figure \ref{th2a:vmap} shows the flux intensity maps of the H$\alpha$ emission line on a logarithmic scale observed in a sequence of 15 velocity channels with a resolution of $\sim 20$\,km\,s$^{-1}$, which can be used to identify different kinematic components of the nebula. The systemic velocity $v_{\rm sys}=-52$\,km\,s$^{-1}$ has been subtracted from the central velocity value given at the top of each channel. The stellar continuum map has also been subtracted from the flux intensity maps. While a prominent equatorial ring can be clearly seen in the $-30$ and $31$\,km\,s$^{-1}$ channels, a pair of collimated bipolar outflows can be identified in the $-71$ and $72$\,km\,s$^{-1}$ channels. This ring has a radius of $14\arcsec$ and a thickness of $9\arcsec$. Two different velocity components seen in the $-51$, $-30$, $31$ and $51$\,km\,s$^{-1}$ channels are consistent with front and back walls of a toroidal shell expanding with a velocity of $\sim 40$\,km\,s$^{-1}$. If we assume that this ring is a projection of a circle on the sky plane, the velocity channels correspond to a position angle (P.A.) of $-45^{\circ} \pm 5^{\circ}$ measured from the north toward the east in the equatorial coordinate system  (ECS). A brightness discontinuity seen at the central region in both directions, from the $51$ to $72$ ($-51$ to $-71$)\,km\,s$^{-1}$ velocity slices, is related to an environment change of collimated point-symmetric outflows emerging from the dense shell. However, the dimensions of collimated bipolar outflows seen pole-on cannot be precisely determined, but their positions are approximately projected near the central star and inside the ring onto the sky plane. It is seen that the bipolar outflows reach a velocity of $\sim \pm 90$\,km\,s$^{-1}$ at the poles, and have similar brightness in the $-71$ and $72$ (also $-92$ and $93$)\,km\,s$^{-1}$ velocity slices. 

\begin{figure*}
\begin{center}
\includegraphics[width=6.0in]{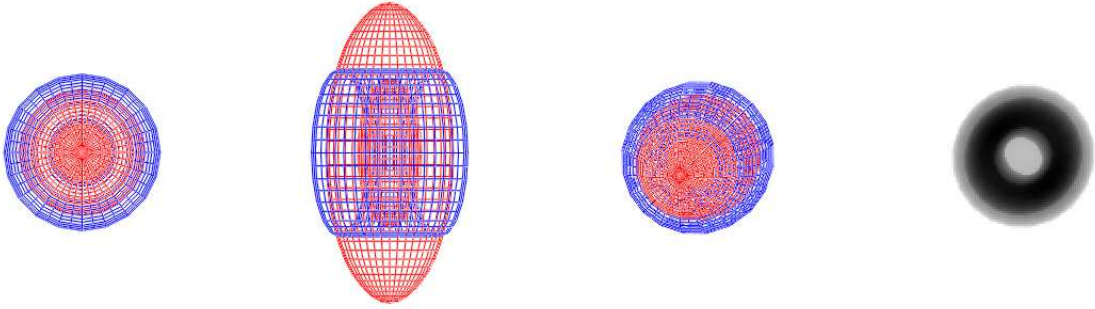}%
\caption{\textsc{shape} mesh model before rendering at two different orientations (inclination: 0$^{\circ}$ and 90$^{\circ}$), the best-fitting inclination, and the corresponding rendered image, respectively.
}%
\label{th2a:shape:model}%
\end{center}
\end{figure*}

\section{Kinematic Modeling}
\label{th2a:sec:models}

To disentangle the three-dimensional gaseous structure of Th\,2-A, we used the morpho-kinematic modeling tool \textsc{shape} \citep[Version 5.0;][]{Steffen2006,Steffen2011}. This program has been used for many objects, such as Hb 5 and K 3-17 \citep{Lopez2012}, NGC 2392 \citep{Garcia-Diaz2012}, Hen 3-1333 and Hen 2-113 \citep{Danehkar2015}, and NGC 3242 \citep{Gomez-Munoz2015}. It uses interactively molded geometrical polygon meshes to reconstruct three-dimensional structures based on kinematic and spatial observations. It constructs a cell grid, each cell representing a volume, and uses a ray-casting algorithm to perform radiative transfer through these cells. It produces several outputs that can be directly compared with observations, namely synthetic images and position--velocity (P--V) diagrams, and velocity channels. However, synthetic images do not include explicit photo-ionization process, so under such conditions the emissivity distribution for each spectral line is modeled ad-hoc based on the observations of the corresponding emission line. To determine a best-fitting model, geometrical and kinematic parameters are iteratively modified until acceptable solutions are obtained.

To model Th\,2-A, we used the H$\alpha$ velocity slices presented in Fig.\,\ref{th2a:vmap} since all the kinematic components are included in H$\alpha$ emission rather [N\,{\sc ii}] emission due to the very high excitation feature of this nebula. The velocity channel maps of the morpho-kinematic model are closely compared with the observed channel maps. All these geometrical structures were slightly modified until the model outputs reasonably match the observational maps. As a starting point, a circular shell was assumed for the ring of Th\,2-A. The velocity is defined as radially outward from the nebular center with a linear function of magnitude, commonly known as a Hubble-type flow \citep{Steffen2009}. The inclination angle of this shell was then manipulated until synthetic images at different velocity channels match the observational maps (see the channels between $-51$ and $51$\,km\,s$^{-1}$). Assuming that the observed circular ring is a toroidal shell, the inclination of major axis is found to be $-10^{\circ} \pm 5^{\circ}$ with respect to the line of sight ($0^{\circ}$ being pole-on, $90^{\circ}$ being edge-on). The best-fitting model describes a toroidal shell with an expansion velocity of $40\pm10$\,km\,s${}^{-1}$.

\begin{table}
\begin{center}
\caption{Parameters of the Best-fitting \textsc{shape} Model of Th\,2-A\label{th2a:parameters}}
\begin{tabular}{lc}
\tableline
\tableline
Parameter & Value \\
\tableline
Inclination of major axis, $i$ & $-10^{\circ} \pm 5^{\circ}$  \\
Position angle of major axis, P.A. 				& $-45^{\circ} \pm 5^{\circ}$  \\
Outer radius of the ring, $r_{\rm out}$			& $14\pm2$ arcsec \\
Thickness of the ring, $\delta r$				& $9\pm2$ arcsec \\
Waist expansion velocity	 	& $40\pm10$\,km\,s${}^{-1}$\\
Polar expansion velocity  				& $90\pm20$\,km\,s${}^{-1}$\\
Systemic velocity (LSR), $V_{\rm sys}$				& $-52\pm5$\,km\,s${}^{-1}$  \\
\tableline
\end{tabular}
\end{center}
\end{table}

\begin{figure*}
\begin{center}
\includegraphics[width=5.5in]{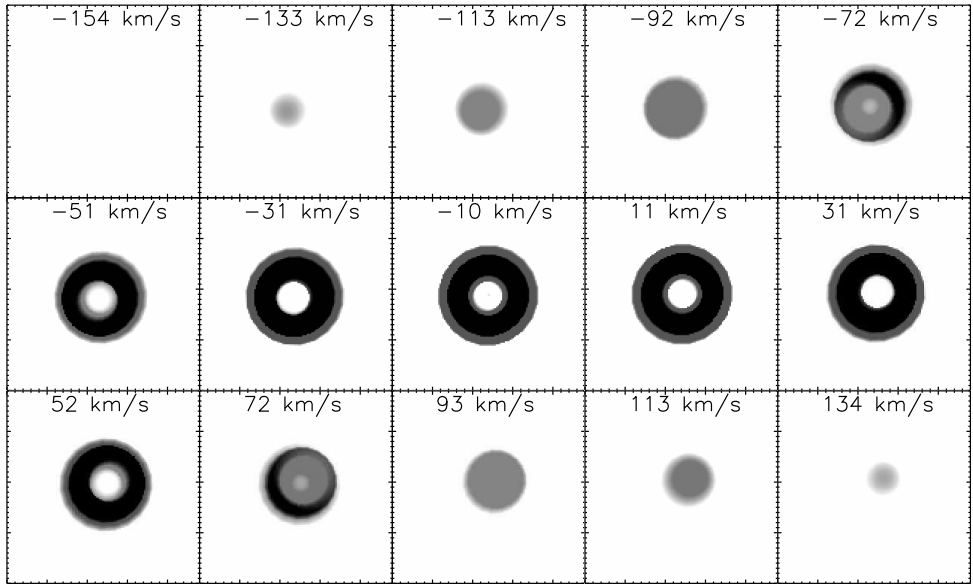}%
\caption{Synthetic images at different velocity channels obtained from the best-fitting \textsc{shape} model. Each channel is 20.5\,km\,s$^{-1}$ wide.
}%
\label{th2a:shape:vmap}%
\end{center}
\end{figure*}

\begin{figure*}
\begin{center}
\includegraphics[width=2.3in]{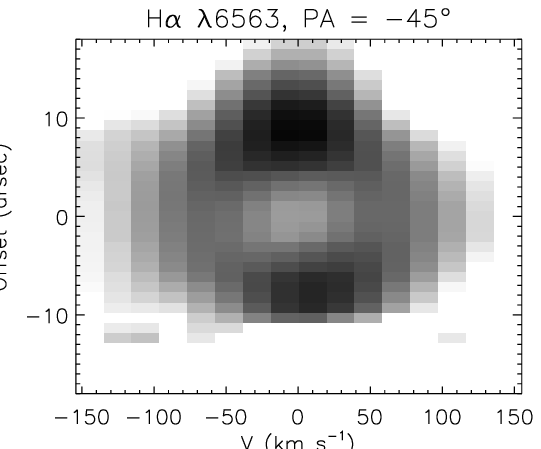}
\includegraphics[width=2.3in]{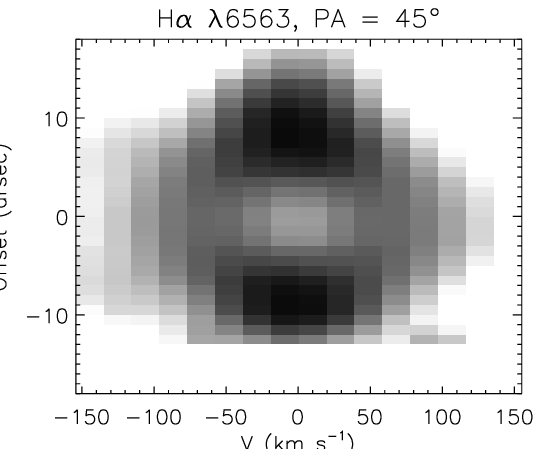}\\~\\
\includegraphics[width=5.0in]{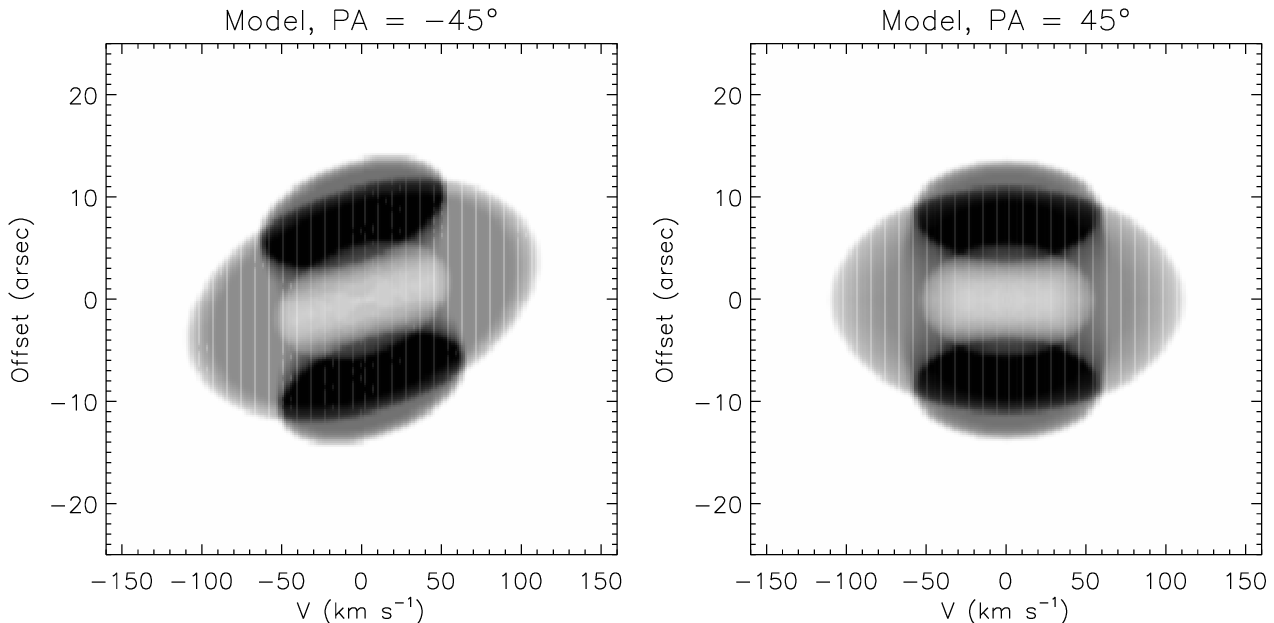}
\caption{Top panels: P--V arrays of Th\,2-A in H$\alpha$ emission for slits oriented with P.A.\,$=-45^{\circ}$ and $45^{\circ}$ passing through the central star. The velocity axis is with respect to to the systemic velocity of the central star, in km\,s${}^{-1}$ unit. The angular offset at 0 arcsec defines the central star position. 
Bottom panels: the corresponding synthetic P--V diagrams obtained from the best-fitting \textsc{shape} model.
\label{th2a:slit}%
}%
\end{center}
\end{figure*}

We further modeled the bipolar collimated outflows using a prolate ellipsoid with a density lower than that of the toroidal shell. It explains the high velocity components seen in the $-92$, $-72$, $72$ and $93$\,km\,s$^{-1}$ channels (see Fig.\,\ref{th2a:vmap}). However, the length of the bipolar outflows cannot precisely be constrained, as they are slightly oriented, relative to the line of sight (inclination $-10^{\circ}$). Assuming a homologous outflow, the distance of the bipolar outflow from the nebular center is nearly twice larger than the polar radius of the toroidal shell. From the model, the collimated bipolar outflows are found to have a polar expansion velocity of $ 90 \pm 20$\,km\,s$^{-1}$.

In Figure \ref{th2a:shape:model}, we present a 3D representation of the final best-fitting model viewed from different orientations (from $0^{\circ}$, $90^{\circ}$), followed by the resultant mesh model (inclination $-10^{\circ}$), before rendering, and the final rendered model, respectively. As seen in Fig.\,\ref{th2a:shape:model}, the obtained synthetic image fairly resembles the observation (see Fig.\,\ref{th2a:fig2}). The parameters of the best-fitting model are summarized in Table~\ref{th2a:parameters}.

The velocity channel maps of the resultant kinematic model are given in Fig.\,\ref{th2a:shape:vmap}, which can be subsequently compared with the observational maps presented in Fig.\,\ref{th2a:vmap}. The close match between them suggests that our morpho-kinematic model is able to reproduce the observed kinematic structure of this object.

Fig. \ref{th2a:slit} (top panels) shows the H$\alpha$ emission line P--V arrays of Th\,2-A extracted from the IFU datacube for two slits oriented with P.A.\,$=-45^{\circ}$ and $45^{\circ}$, which pass through the the central star. We present these slits because the best-fitting model has a symmetric axis with P.A. \,$=-45^{\circ}$. The velocity axis on all plots is relative to the LSR systemic velocity of the central star ($v_{\rm sys}=-52$\,km\,s$^{-1}$). The stellar continuum from the CSPN has also been subtracted. There are two separate velocity components reaching $\pm 110$\,km\,s$^{-1}$. The two bright knots represent the dense torus expanding with a velocity of $40\pm10$\,km\,s$^{-1}$. The synthetic P--V diagrams derived from the model are shown in Fig. \ref{th2a:slit} (bottom panels) under the observed ones. It is seen that they reasonably match the observed diagrams. Note that the south part of Th\,2-A was not completely covered by the IFU field-of-view, which is visible in the lower parts of the P--V arrays.

\section{Conclusions}
\label{th2a:sec:conclusions}

The integral field spectroscopic observations presented in this paper clearly reveal that Th\,2-A, which is known to have a thick ring structure, in fact has also a pair of thin bipolar outflows. We have produced the spatially resolved maps of the H$\alpha$ emission line, including flux intensity and radial velocity (see Fig.\,\ref{th2a:fig2}). In addition, we have derived the flux intensity maps of the H$\alpha$ emission line observed at different velocity channels (see Fig.\,\ref{th2a:vmap}). These observations suggest that Th\,2-A has an equatorial ring structure with a pair of collimated bipolar outflows along its symmetric axis. 

We modeled the observed velocity channels using the 3D interactive tool \textsc{shape}. The geometrical components included in the model are a thick toroidal shell and a thin prolate ellipsoid.
From the reconstruction model, the nebula is found to be tilted by $i=-10^{\circ} \pm 5^{\circ}$ with respect to the line of sight, while its symmetry axis is measured to be P.A.\,$=-45^{\circ} \pm 5^{\circ}$ from the north toward the east in the ECS. It is found that the dense shell has an expansion velocity of $40\pm10$\,km\,s${}^{-1}$, while the collimated bipolar outflows reach an expansion velocity of $90 \pm 20$\,km\,s$^{-1}$.

The high-resolution imaging of the central region of Th\,2-A shows the presence of two stars: CSPN and late-type star \citep{Ciardullo1999,Weidmann2008}. However, a binary system with a separation of $1\farcs 4$ (see Fig. \ref{th2a:hst:fig1}) unlikely contributes to the formation of its point-symmetric outflows. Previously, morpho-kinematic modeling of some PNe around close-binary systems have been shown to have alignments between the nebular shells and the binary orbital inclinations \citep[see e.g.][]{Mitchell2007,Jones2010a,Jones2012,Tyndall2012,Huckvale2013}. A closer inspection of the CSPN Th\,2-A is necessary to examine the existence of an undiscovered companion. A potential triple system could have important implications for its morpho-kinematic structure and rare [WO]-type CSPN. Results of the present study, together with future in-depth studies of its central stars, will help us understand the possible role of close-binary systems in PN morphology. 

\acknowledgments

A.D. acknowledges the support of Research Excellence Scholarship from Macquarie University, and is grateful to Prof.~Quentin~A.~Parker for supporting the ANU\,2.3-m observing run, and the staff at the Siding Spring Observatory for their support. The author also thanks the anonymous referee for constructive comments and suggestions.

Based on observations made with the NASA/ESA \textit{Hubble Space Telescope}, obtained from the Data Archive at the Space Telescope Science Institute, which is operated by the Association of Universities for Research in Astronomy, Inc., under NASA contract NAS 5-26555. 

\textit{Facilities:} \facility{ATT (WiFeS)}, \facility{\textit{HST} (WFPC2)}.

\end{document}